# SUPPRESSION OF THE LONGITUDINAL COUPLED BUNCH INSTABILITY IN DAΦNE IN COLLISIONS WITH A CROSSING ANGLE

Alessandro Drago, Mikhail Zobov, Istituto Nazionale di Fisica Nucleare, Laboratori Nazionali di Frascati, Via Enrico Fermi 40, Frascati, Italy

Dmitry Shatilov, BINP, Novosibirsk, Russia

Pantaleo Raimondi, ESRF, Grenoble, France

*Abstract*

In DAΦNE, the Frascati e+/e- collider operating since 1998, an innovative collision scheme, the crab waist, has been successfully implemented during the years 2008-09. During operations for the Siddharta experiment an unusual synchrotron oscillation damping effect induced by beam-beam collisions has been observed. Indeed, when the longitudinal feedback is off, the positron beam becomes unstable with currents above 200–300 mA due to coupled bunch instability. The longitudinal instability is damped by colliding the positron beam with a high current electron beam (of the order of 2 A). A shift of about -600 Hz in the residual synchrotron sidebands is observed. Precise measurements have been performed by using both a commercial spectrum analyzer and the diagnostic capabilities of the longitudinal bunch-by-bunch feedback. The damping effect has been observed in DAΦNE for the first time during collisions with the crab waist scheme. Our explanation, based both on theoretical consideration and modeling simulation, is that beam collisions with a large crossing angle produce longitudinal tune shift and spread, providing Landau damping of synchrotron oscillations.

## INTRODUCTION

DAΦNE is a Φ-factory, a e+/e- collider built at Frascati in the years 1991-1996 [1], [2] and operating since 1998. DAΦNE accelerator complex is composed by one linac that can accelerate electron beams with energy up to 800 MeV (510 MeV in operation) or positron beams with energy up to 510 MeV, an accumulator-damping ring, a transfer line and two main rings (MR) with one or two interaction points for collisions at 1.02 GeV in the centre of mass. The linear accelerator, the accumulator ring and the transfer line can be set to inject in the MR a single positron or electron bunch every ½ second. In the typical injection scheme of DAΦNE, the electron bunches are stored in the MR before the positron ones because the electron injection efficiency is higher than for the other beam. It must be underlined that the different behaviour is related mainly to the parasitic e-cloud effect limiting the top beam current storable in the e+ ring.

Electrons are therefore stored in MR with a rather high current (≈2A), then the injection system is switched to the positron production mode, operation which takes >1 minute, and, at the end of the switch, the positrons are injected. The electron beam current decays rather rapidly, due to the low beam energy (510 MeV) and to the small transverse emittance, and eventually the two beams collide at approximately the same currents, starting usually in the range between 1.5 and 1.0 A.

After electron injection and during the transfer line switching time, there are beam collisions with very high electron currents (between 2A and 1.5A) and relatively low positron ones (between 500 and 200 mA). In this particular situation, a longitudinal damping of the positron beam has been observed even with the longitudinal bunch-by-bunch e+ feedback turned off. This damping effect has been observed in DAΦNE for the first time during collisions with the crab waist scheme [3], [4], implemented in the 2008 and 2009 years.

After first observations of this behaviour, three dedicated machine study runs have been carried out with the goal of precisely measuring characteristics of the effect [5]. In this paper first of all we describe the measurements results. Then we propose an analytical formula to explain the longitudinal beam-beam tune shift. Finally we compare the measured tune shift with its analytical estimates and numerical calculations.

## INSTRUMENTATION AND MEASUREMENTS

In order to perform the measurements in the positron main ring, we have used two different diagnostic tools comparing the results. Precise measurements on this effect have been performed by using the following systems:

a) a commercial Real-time Spectrum Analyzer RSA 3303 by Tektronix, working from DC to 3 GHz. The spectrum analyzer is connected to a high bandwidth beam pickup, made by four buttons. The bunch signals, after going to the H9, hybrids parts by MA-COM, produce horizontal and vertical differences from the zero orbit and also the sum of the signals.

The sum makes possible the longitudinal oscillation detection. To have a better signal to noise ratio the acquisition system is completed by a bandpass passive filter working at 360MHz and by a small signal amplifier;

b) a longitudinal bunch-by-bunch feedback developed in collaboration with SLAC and LBNL in the years 1993-96 with its powerful beam diagnostic capability both in real time and off line [6], [7], [8]. The feedback has been used in closed loop and open loop.

Fig. 1 shows an image of the spectrum analyzer screen. The highest observed peak at 362.484 MHz corresponds to the 118-th revolution harmonic while the synchrotron sidebands are separated by 35.25 kHz. The e+ feedback is off (i.e. in open loop) and the total positron beam current is 130 mA in 103 bunches.

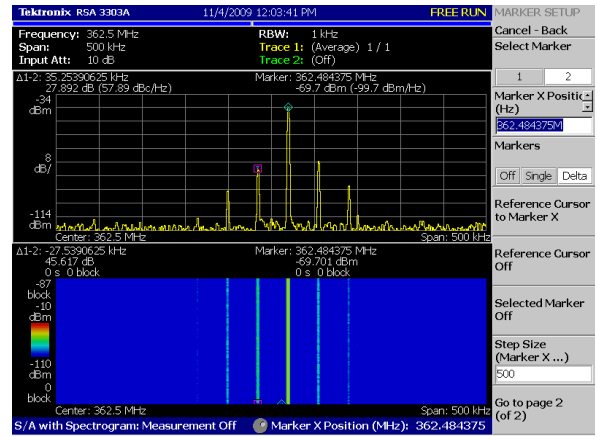

Fig. 1 – Positron beam 118-th revolution harmonic with synchrotron sidebands (feedback off, out of collision)

In the following plot (Fig. 2) showing the positron beam behaviour, the electron beam with ~1700 mA current and the feedbacks on (i.e. in closed loop) is colliding with e+ beam having the longitudinal feedback turned off.

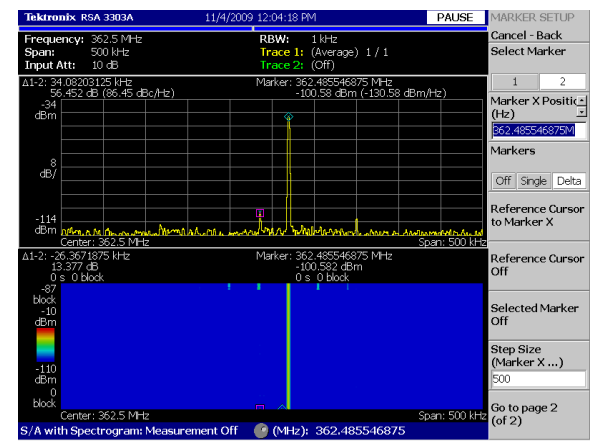

Fig. 2 – Positron synchrotron sidebands damped by beam-beam collisions (feedback off).

The result is clearly visible in the Fig. 2: the e+ beam synchrotron sidebands are almost completely damped when placing the beams in collision, even though the e+ longitudinal feedback is turned off.

Fig. 3 compares the cases "in collision – out of collision" with the same setup as the previous ones, that is with the e+ longitudinal feedback turned off and all the other systems turned on. The sidebands frequency shift of the order of 1kHz is clearly visible but the resolution of the instrument setup is not so accurate to be exactly measured. Nevertheless it is evident from the historical plot that the damping effect induced by the beam-beam collisions makes lower the synchrotron frequency on both sidebands. In the case of Fig. 3, the beam currents are 1550 mA for the electrons and 390 mA for the positrons.

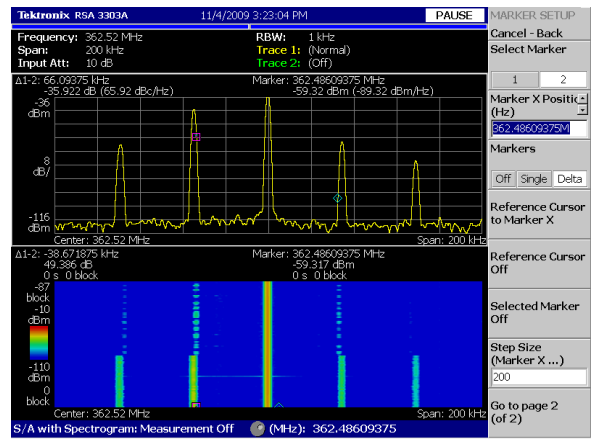

Fig. 3 - Positron beam longitudinal behaviour showing the "in collision – out of collision" cases.

It is also possible to download the traces from RSA 3303A as numerical values. Transferring the data to a PC/MATLAB environment a new plot has been created as shown in the following Fig. 4.

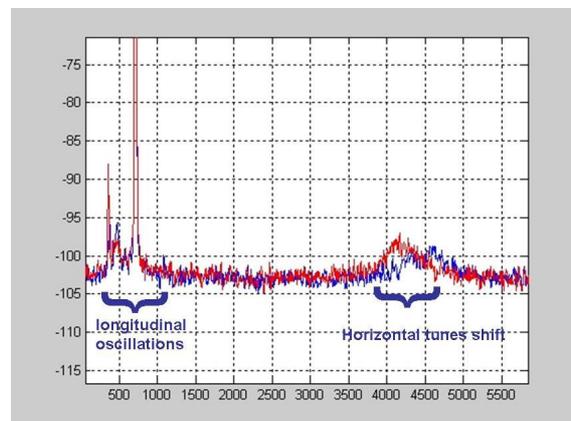

Fig. 4 - Data plot elaborated from the RSA acquisition showing the longitudinal and horizontal tunes in collision (blue) and out of collision (red trace). Vertical scale is in dBm, horizontal axis shows the number of bins (proportional to the frequency)

In Fig. 4 the highest correspons to the e+ 118-th harmonic of the revolution frequency; the red trace shows positron spectrum out of collision, while the blue one corresponds to the case of colliding beams. The vertical scale is in dBm, the horizontal axis is in number of bins (proportional to the frequency). This case is interesting because it shows a situation where the electron beam damps longitudinally and shifts in frequency the positron synchrotron oscillations. Besides, it is possible to see that the beam collisions produce also a horizontal tune shift (in this case increasing the frequency).

With the goal to confirm the measurements done with the spectrum analyzer and to evaluate more precisely the effect, the beam diagnostic tools of the DAΦNE longitudinal feedback have been used. With this system it is possible to record longitudinal data for each bunch. Data can be recorded both in closed loop and in open loop.

Fig. 5 shows the positron beam modal growth rate analysis for the cases respectively in and out of collisions.

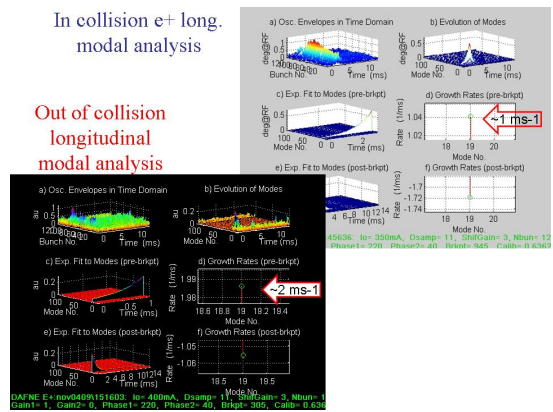

Fig. 5 - Mode 19 growth rate out of collision is 1.99 ms-1. Mode 19 growth rate in collision is 1.04 ms-1

In both cases the mode 19 is the strongest unstable longitudinal mode; out of collisions it has a growth rate, in inverse units, of 1.99 ms-1, (corresponding to 502 microseconds), in collision the growth rate is halved, 1.04 ms-1, corresponding to 961 microseconds. This, once again, confirms the damping effect of beam-beam interaction.

Analyzing with great detail these data, it is possible to measure the synchrotron frequency shift induced on the positron beam by the beam-beam collisions with the e+ longitudinal feedback turned off. As shown in Fig. 6 and 7, the synchrotron frequency (out of collisions) is 34.86 kHz, while the synchrotron frequency (in collisions) is 34.23 kHz. The longitudinal frequency shift induced by the beam-beam collisions is therefore -630 Hz at e+ beam current of 320 mA (out) and 250 mA (in) for the positrons. The e- beam currents were 520 mA (out) and 476 mA (in) respectively.

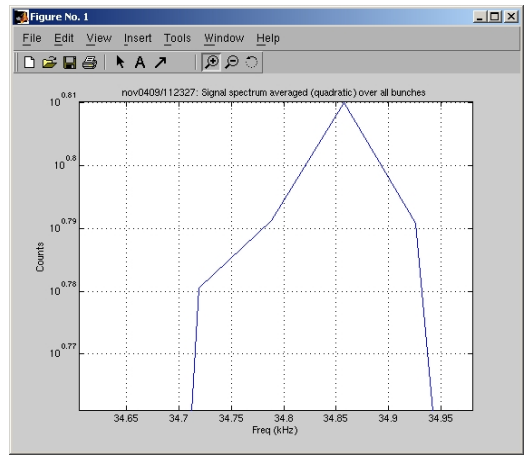

Fig.6 - The e+ synchrotron frequency out of collision is 34.86 kHz.

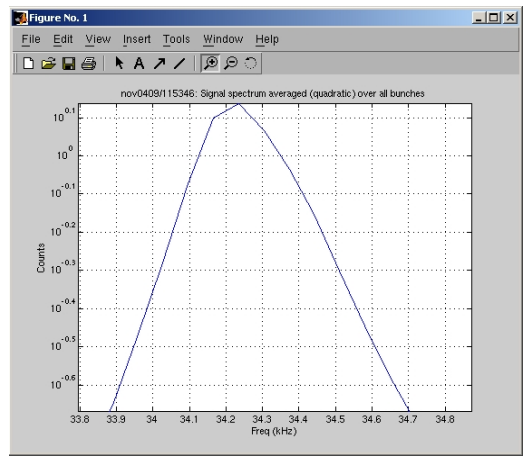

Fig.7 - The e+ synchrotron frequency in collision is 34.23kHz

In the following section, we want to show an analytical expression for the synchrotron tune shift, that is also a measure of the synchrotron tune spread, and eventually to compare the formula with numerical simulations.

## ANALYTICAL FORMULA AND COMPARISON WITH NUMERICAL SIMULATIONS

Summarizing, the experimental observations and measurements at DAΦNE have shown that beam-beam collisions can damp the longitudinal coupled bunch instability. Bringing into collisions a high current electron beam with an unstable positron one was stabilizing the synchrotron oscillations of the e+ beam, even with the longitudinal feedback system switched off. Besides, a negative frequency shift of positron beam synchrotron sidebands has been observed during beam collisions.

The authors attribute these two effects to a nonlinear longitudinal kick arising by beam-beam interaction under a finite crossing angle. It is worthwhile to note here that we have observed this effect clearly only after implementation of the crab waist scheme of beam-beam collisions at DAΦNE having twice larger horizontal crossing angle with respect to the previous operations with the standard collision scheme [9].

In the following an analytical expression for the synchrotron tune shift is outlined [10], and this expression gives also a measure of the synchrotron tune spread. The formula is compared with numerical simulations too.

### Tune shift analytical formula

In collisions with a crossing angle the longitudinal kick of a test particle is created due to a projection of the transverse electromagnetic fields of the opposite beam onto the longitudinal axis of the particle. The kicks that the test particle receives while passing the strong beam with rms sizes $\sigma_x$, $\sigma_y$, $\sigma_z$ under a horizontal crossing angle $\theta$ are [11]:

$$x' = \frac{2r_e N}{\gamma}(x - ztg(\theta/2))\int_0^\infty dw \frac{\exp\left\{-\frac{(x - ztg(\theta/2))^2}{(2(\sigma_x^2 + \sigma_z^2 tg^2(\theta/2)) + w)} - \frac{y^2}{(2\sigma_y^2 + w)}\right\}}{(2(\sigma_x^2 + \sigma_z^2 tg^2(\theta/2)) + w)^{3/2}(2\sigma_y^2 + w)^{1/2}}$$

$$y' = \frac{2r_e N}{\gamma} y \int_0^\infty dw \frac{\exp\left\{-\frac{(x - ztg(\theta/2))^2}{(2(\sigma_x^2 + \sigma_z^2 tg^2(\theta/2)) + w)} - \frac{y^2}{(2\sigma_y^2 + w)}\right\}}{(2(\sigma_x^2 + \sigma_z^2 tg^2(\theta/2)) + w)^{1/2}(2\sigma_y^2 + w)^{3/2}} \quad (1)$$

$$z' = x' tg(\theta/2)$$

where x, y, z are the horizontal, vertical and longitudinal deviations from the synchronous particle travelling on-axis, respectively. N is the number of particles in the strong bunch, $\gamma$ is the relativistic factor of the weak beam. Then, for the on-axis test particle (x = y = 0) the longitudinal kick is given by:

$$z' = -\frac{2r_e N}{\gamma} ztg^2(\theta/2) \int_0^\infty dw \frac{\exp\left\{-\frac{(ztg(\theta/2))^2}{(2(\sigma_x^2 + \sigma_z^2 tg^2(\theta/2)) + w)}\right\}}{(2(\sigma_x^2 + \sigma_z^2 tg^2(\theta/2)) + w)^{1/2}(2\sigma_y^2 + w)^{1/2}} \quad (2)$$

For small synchrotron oscillations $z \ll \sigma_z$ the exponential factor in the integral can be approximated by 1 and we obtain an expression for the linearized longitudinal kick:

$$z' = -\frac{2r_e N}{\gamma} ztg^2(\theta/2) \frac{1}{(\sigma_x^2 + \sigma_z^2 tg^2(\theta/2)) + \sqrt{(\sigma_x^2 + \sigma_z^2 tg^2(\theta/2))\sigma_y^2}} \quad (3)$$

For the case of flat beams with $\left(\sigma_y \ll \sqrt{\sigma_x^2 + \sigma_z^2 tg^2(\theta/2)}\right)$ the tune shift expression can be simplified to [10]

$$\xi_z = -\frac{r_e N^{strong}}{2\pi\gamma^{weak}} \frac{\left(\frac{\sigma_{z0}}{(\sigma_E/E)}\right)^{weak}}{\left(\left(\frac{\sigma_x}{tg(\theta/2)}\right)^2 + \sigma_z^2\right)^{strong}} \quad (4)$$

As we see from (4), for the flat bunches the synchrotron tune shift practically does not depend on the vertical beam parameters. So, one should not expect any big variations due to crabbing and/or hour-glass effect. Since particles with very large synchrotron amplitudes practically do not "see" the opposite beam (except for a small fraction of synchrotron period) their synchrotron frequencies remain very close to the unperturbed value $\nu_{z0}$. For this reason, like in the transverse cases, the linear tune shift can be used as a measure of the nonlinear tune spread.

### Numerical Simulations

In order to check validity of the formulae we performed numerical simulations with the beam-beam code LIFETRAC [12] comparing the tune shift calculated numerically with the one obtained by using the analytical formula (4). As it has been shown [10] by using the typical parameters of SuperB [13] and DAFNE, the formula agrees well with the simulations when the horizontal tune is far from the linear synchrobetatron coupling resonance, see Fig. 8. The agreement improves for larger Piwinski angles.

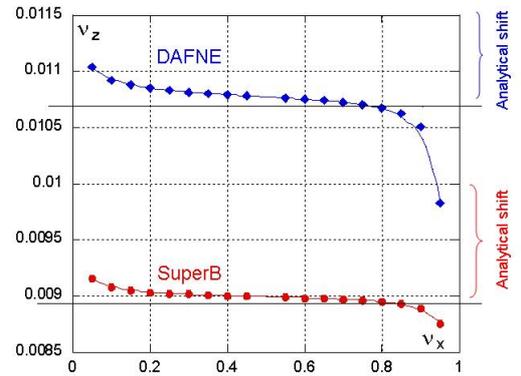

fig. 8 - Synchrotron tune dependence on the horizontal tune. The solid straight lines correspond to the analytically predicted synchrotron tunes

In Fig. 9 the blue curve shows the calculated synchrotron tune dependence on the normalized synchrotron amplitude for the DAΦNE "weak" positron beam interacting with the "strong" electron beam having a current of 1.7 A. For comparison, the green curve shows the tune dependence on amplitude arising due to nonlinearity of the RF voltage. As we can see, the synchrotron tune spread due to the beam-beam interaction is notably larger than that due to the RF voltage alone, at least within 5 $\sigma_z$. In the past it was shown that the RF voltage nonlinearity is strong enough to damp quadrupole longitudinal couple bunch mode instability [14]. So, we can expect a strong Landau damping of longitudinal coupled bunch oscillations by the beam-beam collision. This conclusion is in accordance with performed measurements.

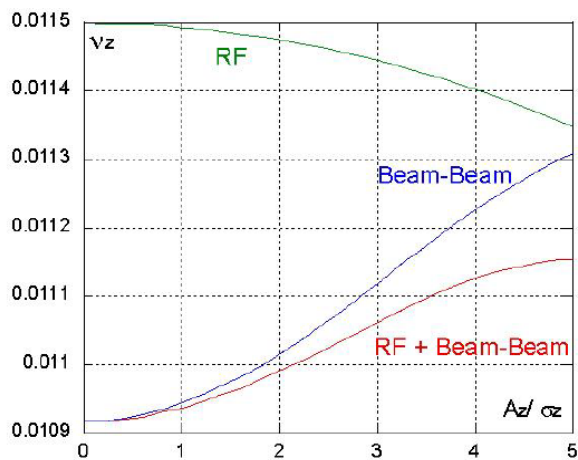

Fig. 9 - Synchrotron tune dependence on normalized amplitude of synchrotron oscillations (blue curve – tune dependence created by beam-beam collisions alone, green – RF nonlinearity alone, red – both contributions).

Summarizing the results of the computation, first of all our numerical simulations have confirmed that the synchrotron tune shift does not depend on parameters of the vertical motion. As a second point, the agreement between the analytical and numerical estimates is quite reasonable for the horizontal tunes far from integers. Quite naturally, in a scheme with a horizontal crossing angle, synchrotron oscillations are coupled with the horizontal betatron oscillations. One of the coupling's side effects is the $\nu_z$ dependence on $\nu_x$, which becomes stronger in vicinity of the main coupling resonances. In order to make comparisons with the analytical formula we need to choose the horizontal betatron tune $\nu_x$ closer to half-integer, where its influence on $\nu_z$ is weaker. The coupling vanishes for very large Piwinski angles. Since $\nu_x$ for DAΦNE is rather close to the coupling resonance, we use numerical simulations in order to compare the calculated synchrotron tune shift with the measured one. In particular, when colliding the weak positron beam with 500mA electron beam, the measured synchrotron frequency shift was about -630 Hz (peak-to-peak). In our simulations we use the DAΦNE beam parameters with respectively bunch current $N=0.9 \times 10^{10}$ and bunch length $\sigma_z = 1.6$ cm. These values give a result in the synchrotron tune shift of -0.000232 corresponding to the frequency shift of -720 Hz. In our opinion the agreement is good considering experimental measurement errors and the finite width of the synchrotron sidebands.

## CONCLUSION

The synchrotron oscillations damping in beam-beam collisions with a crossing angle has been observed in DAΦNE [15], [16]. The respective experimental data have been collected by the commercial spectrum analyzer and by the bunch-by-bunch longitudinal feedback system. The measurement results obtained by the two diagnostics tools are in a good agreement. A simple analytical formula to explain synchrotron tune shift and tune spread due to beam-beam collisions with a crossing angle has been presented. The formula agrees well with the simulations when the horizontal tune is far from the synchro-betatron resonances. The agreement is better for larger Piwinski angles. Calculations have shown that at high beam currents the synchrotron tune spread induced by the beam-beam interaction at DAΦNE can be larger than the tune spread due to the nonlinearity of the RF voltage. This may result in additional Landau damping of the longitudinal coupled bunch oscillations. The effect of the longitudinal kick arising in collisions with a crossing angle has been taken into account in design of the low energy electron-positron collider for production and study of (μ+μ-) bound state in Novosibirsk [17] as well as for the evaluation of the energy change at the IP in the FCC-ee [18].

## REFERENCES

[1] The DAΦNE Project Team presented by G.Vignola, " DAΦNE, The Frascati PHI-factory". PAC '93, Washington, May 1993, p.1993.

[2] The DAΦNE Project Team presented by G.Vignola, "DAΦNE Status and Plans", 16th IEEE Particle Accelerator Conference (PAC 95), Dallas, Texas, May 1-5, 1995. Published in IEEE PAC 1995: 495-499.

[3] P. Raimondi et al., "Beam-Beam Issues for Colliding Schemes with Large Piwinski Angle and Crabbed Waist", e-Print:physics/0702033, 2007.

[4] M. Zobov et al., "Test of crab-waist collisions at DAΦNE Phi factory", Phys.Rev.Lett. 104 (2010) 174801.

[5] A. Drago, P. Raimondi and M. Zobov, "Beam-Beam Longitudinal Damping", DAΦNE Technical Note G-71, 08 February 2010.

[6] J. Fox et al., "Observation, Control and Modal Analysis of Coupled-Bunch Longitudinal Instabilities", EPAC'96 Proceedings, 1996.

[7] D. Teytelman, "Architectures and Algorithms for Control and Diagnostics of Coupled-Bunch Instabilities in Circular Accelerators". Stanford Univ. PhD thesis, SLAC-R-633, June 2003.

[8] S. Prabhakar, "New Diagnostics and Cures for Coupled-Bunch Instabilities", Stanford Univ. PhD thesis, SLAC-R-554, August 2001.

[9] C. Milardi et al., "Crab Waist Collision at DAΦNE", published in ICFA Beam Dyn. Newslett. 48, pp. 23-33, 2009.

[10] M. Zobov and D. Shatilov, "Synchrotron Tune Shift and Tune Spread Due to Beam-Beam Collisions with a Crossing Angle", DAΦNE Technical Note G-72, 04 March 2010.

[11] P. Raimondi and M. Zobov, "Tune Shift in Beam-Beam Collisions with a Crossing Angle", DAΦNE Technical Note G-58, 26 April 2003.

[12] D. Shatilov, "Beam-beam simulations at large amplitudes and lifetime determination", Part. Accel., vol.52 , pp. 65-93, Jul. 1996.

[13] "SuperB: A High-Luminosity Asymmetric e+ e- Super Flavor Factory. Conceptual Design Report", e-Print: arXiv:0709.0451 [hep-ex].

[14] M. Migliorati *et al.*, "Landau Damping of Longitudinal Multi-Bunch Instabilities in DAΦNE", DAΦNE Technical Note G-21, 13 September 1993.

[15] A. Drago *et al.*, "Synchrotron Oscillation Damping Due to Beam-Beam Collisions", THOBRA01, oral talk presented at the first International Particle Accelerator Conference held in Kyoto, Japan, May, 23-28, 2010.

[16] A. Drago, P. Raimondi, D. Shatilov, and M. Zobov, "Synchrotron oscillation damping by beam-beam collisions in DAΦNE", Phys. Rev. ST Accel. Beams, Vol. 14, p. 092803, Sep. 2011. doi: 10.1103/PhysRevSTAB.14.092803

[17] A.Bogomyagkov *et al.*, "Low-energy electron-positron collider to search and study (μ+μ-) bound state", EPJ Web of Conferences 181, 01032 (2018). doi.org/10.1051/epjconf/201818101032

[18] D. Shatilov, "Impact of beam-beam effects on beam energy and crossing angle at IP", talk at FCC Week 2019, 24-28 June 2019, Brussels, Belgium.